\documentclass[10pt,twocolumn]{article}

\usepackage{graphicx,amsmath,amssymb,textcomp,authblk,upgreek,fullpage}
\usepackage[utf8x]{inputenc}
\date{\vspace{-8ex}}
\author[1]{David J. French}
\author[1]{Andrew B. Schofield}
\author[1,*]{Job H.J. Thijssen}
\affil[1]{SUPA, School of Physics \& Astronomy, The University of Edinburgh, Peter Guthrie Tait Road, Edinburgh, EH9 3FD, UK}
\affil[*]{j.h.j.thijssen@ed.ac.uk}
\title{Bicontinuous soft solids with a gradient in channel size designed for energy storage applications}
\begin{document}
\twocolumn[
\begin{@twocolumnfalse}
\maketitle

\begin{abstract}
\noindent We present examples of bicontinuous interfacially jammed emulsion gels (``bijels'') with a designed gradient in the channel size along the sample. These samples are created by quenching binary fluids which have a gradient in particle concentration along the sample, since the channel size is determined by the local particle concentration. A gradient in local particle concentration is achieved using a two-stage loading process, with different particle volume fractions in each stage. Confocal microscopy and image analysis were used to quantitatively measure the channel size of the bijels. Bijels with a gradient in channel size of up to 2.8\%/mm have been created.  Such tailored soft materials could act as templates for energy materials optimised for both high ionic transport rates (high power) and high interfacial area (high energy density), potentially making them useful in novel energy applications.
\end{abstract}
\end{@twocolumnfalse}
]
\section{Introduction}
Bijels, which are bicontinuous emulsions stabilised by solid colloidal particles, were first predicted in simulations in 2005 \cite{Stratford_05}, and were first realised experimentally in 2007 \cite{Herzig_07}. Bijels consist of two inter-penetrating fluid channels, with channel sizes typically around $10~\upmu$m, and a layer of colloidal particles at the interface between the two fluids. The particle layer is jammed, preventing the liquid channels from coarsening, yielding a kinetically stable structure with a large interfacial area between the two fluids (even though the particles occupy $\approx$ 90\% of the interface, the remaining interstitial area is still large, on the order of 1~m$^2$ per mL of sample) \cite{Cates_08,Mohraz_16,Welch_17}. 

The bicontinuous structure and large interfacial-area-to-volume ratio of bijels make them promising candidates for use in novel energy materials, especially since they can be (post)-processed to allow the initial fluid channels to be replaced with other materials \cite{Lee_10,Lee_13,CleggChap2,Cordoba_19,Garcia_19,McDevitt_19a}. For example, Lee {\it et al.} have used water-lutidine bijels as scaffolds to synthesise porous gold monoliths \cite{Lee_14}, while Mohraz {\it et al.} have developed three-dimensional Ni/Ni(OH)$_2$ composite electrodes \cite{Witt_16} and ZnO electrodes \cite{McDevitt_19b}. Similarly, Zekoll {\it et al.}
have fabricated hybrid electrolytes composed of bicontinuous microchannels of ceramic lithium-ion electrolyte and a non-conducting polymer \cite{Zekoll_18}, while Cai {\it et al.} have developed polystyrene-filled bicontinuous structures using ethylene carbonate/xylene bijel templates \cite{Cai_18}. Recently, Ching {\it et al.} demonstrated a method using partially miscible mixtures of solvent and poly(ethylene glycol) to create macroporous bijel-templated materials in timescales of minutes rather than hours \cite{Ching_21}.

Whilst the large interfacial area of bijels makes them a promising candidate for novel energy materials, increasing that interfacial area by reducing the channel size also results in a reduced rate of diffusion through the sample \cite{Mezedur_02}, limiting the electrodes' performance \cite{Harmas_18}. To try and optimise both the interfacial area and the diffusive transport rate, we have designed and developed bijels which have a gradient in channel size - allowing material to diffuse quickly into the sample through the large channels at one end of the sample, whilst maintaining a large interfacial area at the other end of the sample. This structure is similar to that of blood vessels, where a gradient in channel width, from arteries to capillaries, optimises the rates at which blood is transported around the body and at which substances transfer between blood vessels and surrounding cells \cite{Costanzo}.

We create bijels using two fluids (ethylene glycol and nitromethane) which are not miscible at room temperature, but which become miscible above $\approx$~45~\textcelsius~\cite{Tavacoli_11}. When the sample is mixed together above this critical temperature and subsequently quenched through the critical point (by immersing the sample in room temperature water), the two liquids separate by spinodal decomposition \cite{Stratford_05}. We ensure separation by spinodal decomposition rather than nucleation by fixing the mass ratio of the two liquids at 64:36 (nitromethane:ethylene glycol), such that the quench goes through the critical point \cite{Witt_13,Lee_13}. However, the inclusion of neutrally wetting colloidal silica particles in the mixture arrests this separation, as the particles adsorb to the liquid-liquid interface and jam when the liquids' interfacial area is approximately equal to the total cross-sectional area of the particles. The channel size in a bijel, $L$, is therefore related to the particle radius, $r_p$, and the particle concentration, $\phi_p$, by \cite{Cates_08,Welch_17}:

\begin{equation}
 L \propto \frac{r_p}{\phi_p}.
\end{equation}
 
\begin{figure}[ht]
 \begin{center}
\includegraphics[width=0.95\columnwidth]{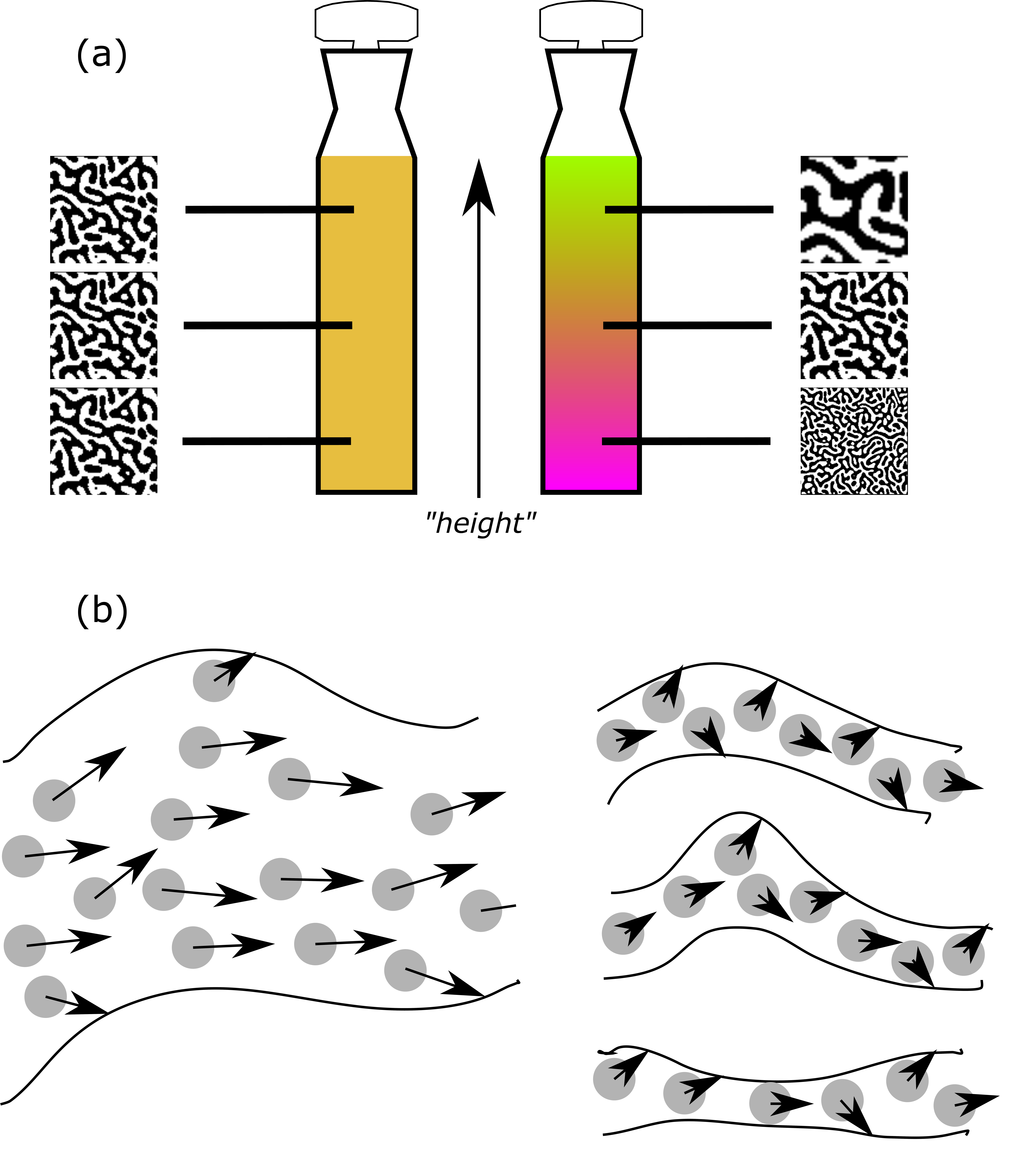}
\caption{(a) Schematics of bijel samples without (left) and with (right) gradients in channel size. (b) Schematics showing how a gradient in channel size could help optimise transport rate and interfacial area. When the channel size is large (left), diffusion of reactant (grey spheres) along the channel is relatively quick, but the relative interfacial area between channels is low, limiting the reaction rate at the interface. When the channel size is small (right), a greater reaction rate is possible due to more interfacial area being available per unit volume, but the movement of reactants along the channel is reduced.}
  \label{fig:IntroFig}
 \end{center}
\end{figure}

We create our samples by quenching binary fluids which have a gradient in particle concentration. Since the timescale of fluid-fluid separation is much shorter than that of particle diffusion \cite{Kim_08}, when such samples are quenched, the local channel size is determined by the local particle concentration, creating a bijel with a gradient in channel size. Figure~\ref{fig:IntroFig}~(a) shows schematics comparing a bijel with a designed gradient in channel size to a conventional uniform bijel and Fig.~\ref{fig:IntroFig}~(b) shows how a bijel with a gradient in channel size could optimise the diffusion rate and interfacial area of a sample. In this case, the channel size becomes a function of height in the sample, $z$:

\begin{equation}
 L(z) \propto \frac{r_p}{\phi_p(z)}.
 \label{Eqn:LocalPhiEqn}
\end{equation}

Bijels conventionally have had channels which are approximately uniform in size across the sample \cite{CleggChap1}, although Haase {\it et al.} have produced narrow bijel fibres with a gradient in channel size due to solvent transfer-induced phase separation \cite{Haase_15,Boakye-Ansah_19}. In this paper we present the results of experiments that produce bijels, through temperature-induced phase separation, with a designed gradient in channel size, which may be useful for optimising bijel-templated electrodes to provide both high power and energy densities. In the work presented below, we purposefully create a gradient in the particle concentration using a two-stage loading process. When such samples are quenched, the local particle concentration determines the local channel size, creating a bijel with a gradient in channel size. We foresee such samples being useful as templates for novel energy storage devices, where both high power and high energy density are desirable. This will be particularly useful if combined with novel techniques which have the potential to scale up the volume of bijels produced, such as direct mixing/homogenisation \cite{Cai_17,Huang_17} and phase inversion \cite{Li_20}.

\section{Materials and Methods}
\subsection{Materials}
Monodisperse, spherical Rhodamine B-labelled fluorescent silica particles were synthesised by following a modified St\"ober method \cite{Stober_68,vanBlaaderen_92}. The silica particles were washed ten times with distilled water, and were then hydrophobised using hexamethyldisilazane (HMDS) (Aldrich, 99.9\%) in a solution of ethanol (Fisher, 99.8\%) and ammonia (Fisher, 35\%) \cite{Tavacoli_11,Hijnen_15,Capel-Sanchez_04,Suratwala_03}. The amount of HMDS used is varied to ensure that the particles are neutrally wetting at the fluid-fluid interface (non-neutrally wetting particles result in an emulsion being formed \cite{Tavacoli_11, White_11,Jansen_11}). Prior to being used to form bijels, the particles were dried in an oven (Binder VD 23) at 170~\textcelsius\ for 30 minutes at atmospheric pressure before being ground up using a spatula and dried for a further 60 minutes at 170~\textcelsius\ under full vacuum. The average particle diameter and polydispersity were calculated as 600~nm and 20\% by measuring 200 individual particles in a scanning electron microscope image. For the purposes of calculating particle volume fractions the particle density is assumed to be 1750~kg\,m$^{-3}$ \cite{vanBlaaderen_92}. The size and shape of the particles was checked qualitatively using confocal microscopy.

Nitromethane (Acros Organics, 99+\%), ethylene glycol (Sigma-Aldrich, 99.8\%) and fluorescein acid (Aldrich) were used as supplied.

\subsection{Methods}

\subsubsection{Bijel fabrication approach}
\begin{figure}[ht]
 \begin{center}
\includegraphics[width=0.9\columnwidth]{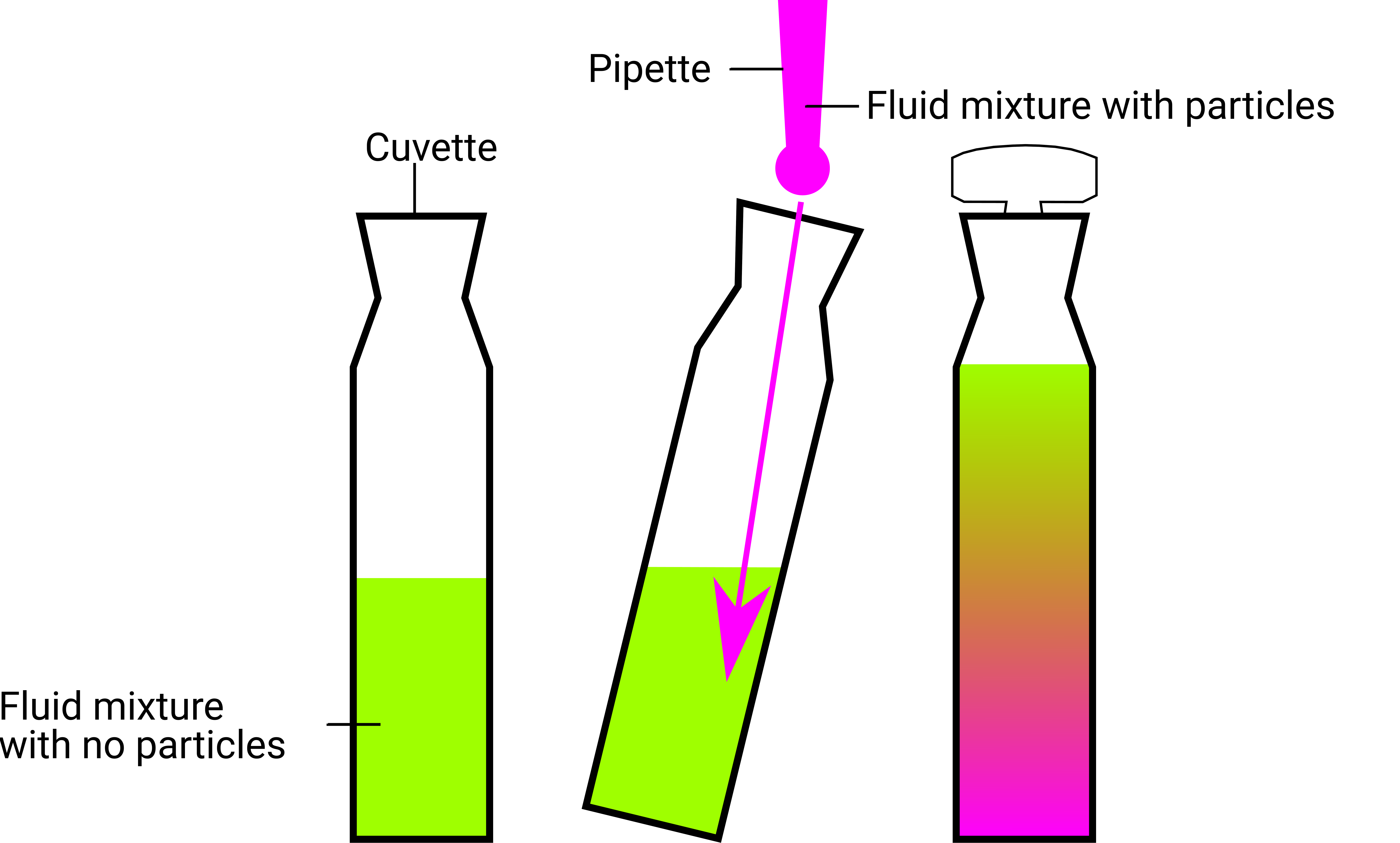}
\caption{Schematic showing preparation of non-uniform bijel. Left: The pre-heated cuvette is initially partially filled with a mixture of nitromethane and ethylene glycol which contains no particles (green). Centre: A mixture of nitromethane and ethylene glycol which does contain particles (red) is gently added to the cuvette using a pipette, so that it runs down the edge of the cuvette. Right: This creates a dispersion with a gradient in particle concentration which, when quenched, yields a bijel with a gradient in channel size.}
  \label{fig:GradedBijelMethodSchem}
 \end{center}
\end{figure}

The appropriate amount of dried silica particles was placed in a 7 mL glass vial, and the appropriate masses of nitromethane and (fluorescein-labelled) ethylene glycol were added. A typical sample would contain 0.03~g of silica, 0.64~g of nitromethane and 0.36~g of ethylene glycol. The vial was then placed in an ultrasonic bath (VWR USC300T) for approximately 10 minutes to disperse the particles. Hot water was used in the bath so that the two fluids also became miscible whilst the particles dispersed. Following dispersal, the vial was placed in an aluminium block which had been pre-heated to 70~\textcelsius. To produce uniform bijels, a pre-heated glass pipette was then used to transfer the sample into a pre-heated glass cuvette (Starna type 21 glass cell, 1~mm path length), which was quickly quenched by submersing it in cold water (19~\textcelsius; room temperature was 24~\textcelsius). To produce non-uniform bijels, the pre-heated cuvette is first (approximately) half-filled with a mixture of nitromethane and ethylene glycol (nitromethane:ethylene glycol mass ratio = 64:36; $\phi_p=0$), before the particle dispersion (nitromethane:ethylene glycol mass ratio = 64:36; $\phi_p\approx3\%$) is gently added to the cuvette. When the dispersion is added, it runs down the inside edge of the cuvette, mixing with the fluid that is already in the cuvette to form a dispersion with a gradient in particle concentration (high at the bottom of the cuvette, low at the top). A schematic of this process is shown in Fig.~\ref{fig:GradedBijelMethodSchem}.

\begin{figure*}[ht]
	\begin{center}
		\includegraphics[width=0.75\textwidth]{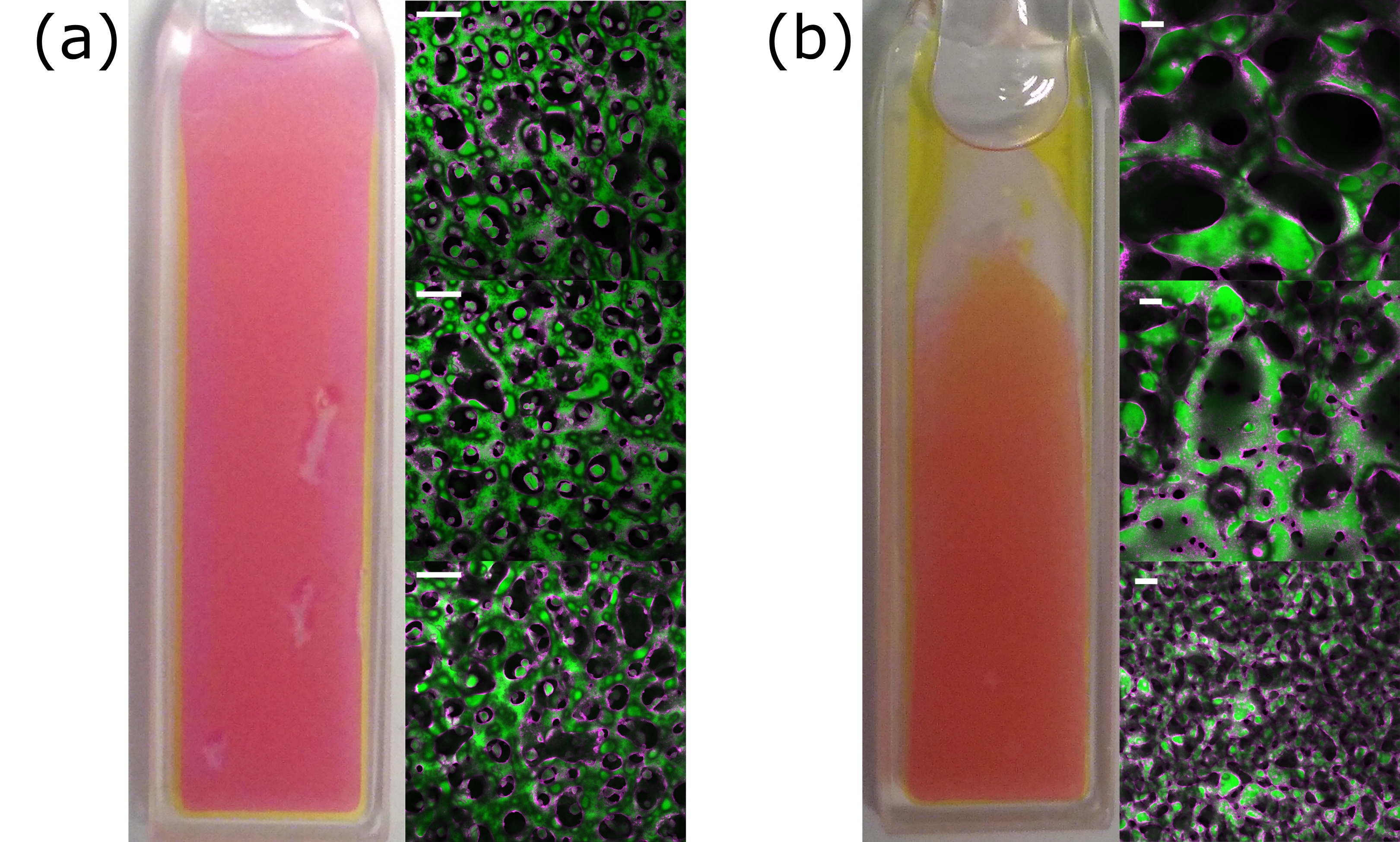}
		\caption{(a): A bijel sample with no gradient in channel size. (b) A bijel sample with a gradient in channel size. In the photographs the cuvettes are 1.25 cm wide; the scale bars in the confocal images are 100 $\upmu$m. In the confocal images, the green channel is the fluorescein-labelled ethylene glycol, the magenta channel is the Rhodamine B-labelled silica and the black is the nitromethane channel.}
		\label{fig:ResultsFig1}
	\end{center}
\end{figure*}

\subsubsection{Characterisation and image analysis}
A Zeiss LSM 700 confocal microscope, coupled to a Zeiss Axio Observer Z1 inverted microscope, was used to perform confocal microscopy. A 10~mW max, 488~nm solid-state laser was used to excite the fluorescein-labelled ethylene glycol and a 10~mW max, 555~nm solid-state laser was used to excite the Rhodamine B-labelled silica particles; emission filters were used as appropriate. Imaging was carried out using Zeiss Plan-Neofluar objective lenses: 10$\times$ (0.3NA), 20$\times$ (0.4NA) and 40$\times$ (0.6NA).

A Kr{\"u}ss DSA100 tensiometer was used to take bright field images of cuvettes containing binary fluid mixtures, and of the bijels produced by quenching these mixtures. The Plot Profile command in ImageJ was used to measure the greyscale values in the samples as a function of height in the cuvettes. These greyscale values were used to calculate the transmission through the sample as a function of height, $T(z)$, by normalising by the greyscale values measured for a cuvette filled with a binary fluid with $\phi_p=0$.

The absorbance, $A(z)$, of a sample is related to the local number density, $n(z)$, of particles by the Beer-Lambert law:
\begin{equation}
    T(z)=\exp(-\alpha\, b\, n(z))
    \label{Eqn:BeerLambertEqn}
\end{equation}
\begin{equation}
    A(z)=-\log_{10}T(z)\propto n(z)
    \label{Eqn:AbsorbanceEqn}
\end{equation}
where $\alpha$ is a constant and $b$ is the path length of the sample \cite{Swinehart_62}.

Quantitative image analysis was carried out using GNU Octave \cite{Octave}. The green (ethylene glycol) channel from each confocal image was used, and the following protocol followed. First a Gaussian filter was used to smooth the image slightly, then the image was thresholded so that 48\% of pixels were white and 52\% were black. These fractions were chosen as these are the relative volumes of the ethylene glycol-rich and ethylene glycol-poor phases, respectively \cite{Tavacoli_11}. The fast Fourier transform (FFT) of the thresholded image was then taken. Following Ref.~\cite{Rumble_16}, the radially-averaged FFT was fitted to the sum of two Lorentzian functions:
\begin{equation}
	y=\frac{A}{1+\left(\frac{x}{w_1}\right)^2}+\frac{B}{\left(1+\frac{x}{w_2}\right)^2} + C,
	\label{FitEquation}
\end{equation}
where $y$ is the intensity, $x$ is the distance in the FFT in  $\upmu$m$^{-1}$ and $A$, $B$, $C$, $w_1$ and $w_2$ are fitting parameters. The full width at half maximum (FWHM) is calculated from the fitted equation. The typical lengthscale in the image is taken as the reciprocal of the FWHM.

Statistical analysis of the measured lengthscales was carried out by fitting data to simple linear regressions and calculating the coefficient of determination, $R^2$ , using:
\begin{equation}
 R^2=\frac{\left( \sum\limits_i(x_i-\bar{x})(y_i-\bar{y})\right)^2}{\sum\limits_i(x_i-\bar{x})^2\sum\limits_i(y_i-\bar{y})^2} \ ,
 \label{Eqn:RSquared}
\end{equation}
where $x_i$ and $y_i$ are the co-ordinates for the $i^{\mathrm{th}}$ datapoint, and $\bar{x}$ and $\bar{y}$ are mean values of $x$ and $y$. $R^2$ is also known as the square of the Pearson product-moment correlation coefficient.

\section{Results}

\begin{figure*}[ht]
 \begin{center}
\includegraphics[width=\textwidth]{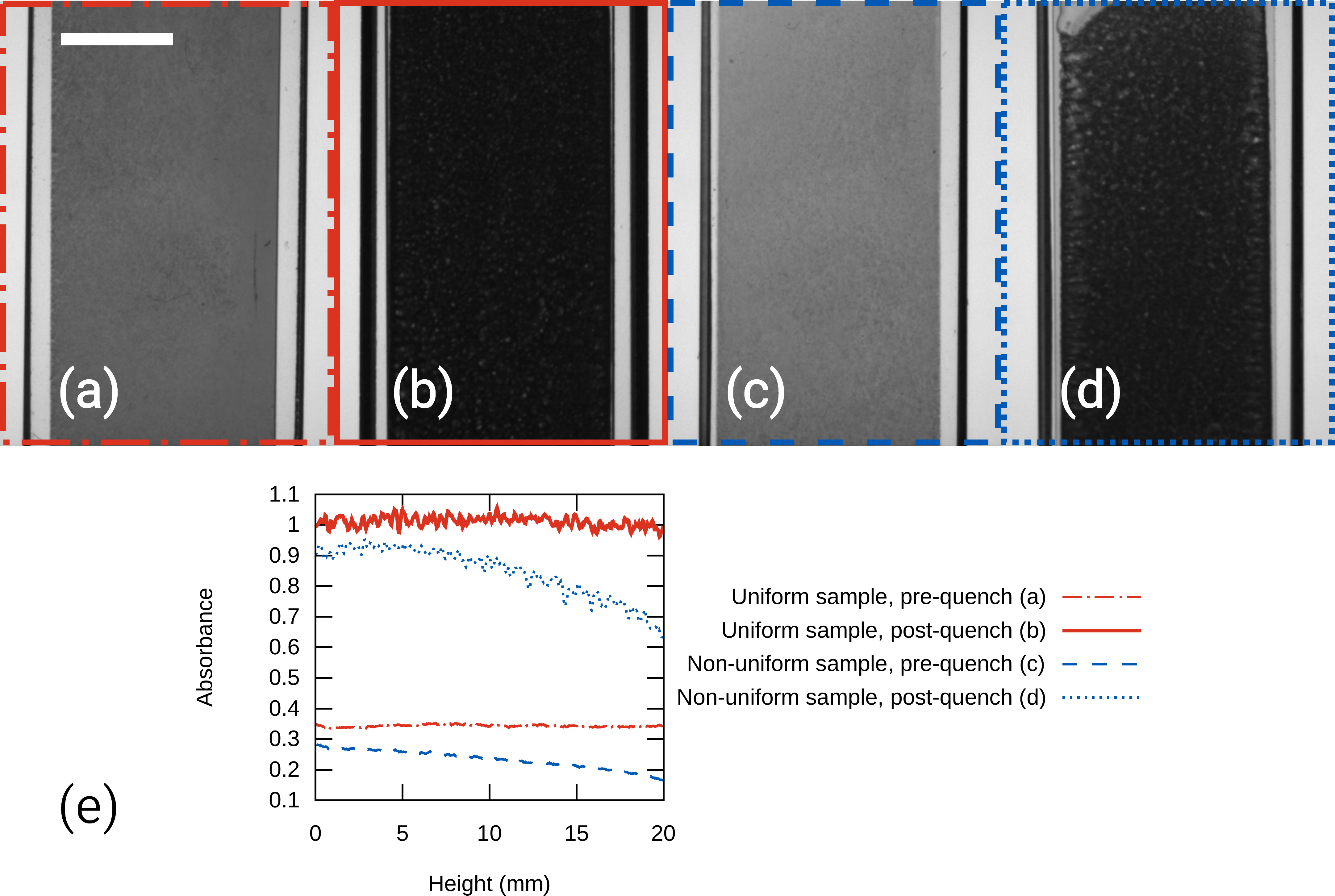}
\caption{Top: Photographs of cuvettes containing uniform and non-uniform bijel mixtures pre- and post-quench. The scale bar is 0.5 cm. Bottom: Graph showing the measured absorbance in the cuvettes as a function of height.}
  \label{fig:TransmissionFig}
 \end{center}
\end{figure*}

Photographs and confocal microscopy images of bijels, with and without gradients in channel size, are shown in Fig.~\ref{fig:ResultsFig1}. Both samples have average particle volume fractions of 2\%. The confocal images in Fig.~\ref{fig:ResultsFig1}~(b) clearly show that the channel size increases with height in the sample. The photograph also shows that the sample is darker at the bottom, due to increased scattering of light at smaller channel sizes and concurrently higher particle concentration. 

The absorbance, $A(z)$, of two bijel samples, one with a uniform channel size and one with a gradient in channel size, is shown in Fig.~\ref{fig:TransmissionFig} (b,d). This shows that the measured absorbance is approximately unchanged along the length of the uniform sample but decreases with sample height in the sample with a gradient in channel size. This is expected, since smaller bijel channels will cause more scattering and hence appear darker in the transmission images. We have also used transmission images to measure the particle volume fraction, $\phi_p(z)$, in the samples prior to quenching (Fig.~\ref{fig:TransmissionFig}, (a,c)). This corroborates the relationship between the pre-quench local particle volume fraction and the post-quench channel widths, as expected from Eqn.~\ref{Eqn:LocalPhiEqn}. Figure~\ref{fig:TransmissionFig} (e) shows how the absorbance in each of the samples changes as a function of height in the cuvette.

\begin{figure*}[ht]
 \begin{center}
\includegraphics[width=0.49\textwidth]{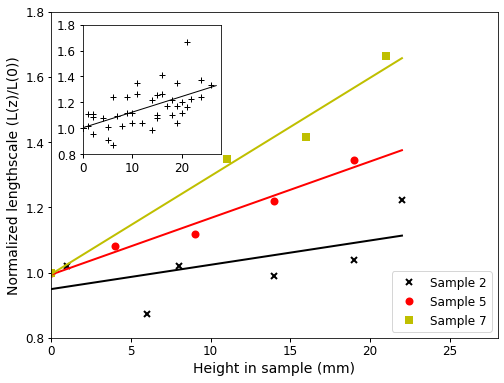}
\includegraphics[width=0.49\textwidth]{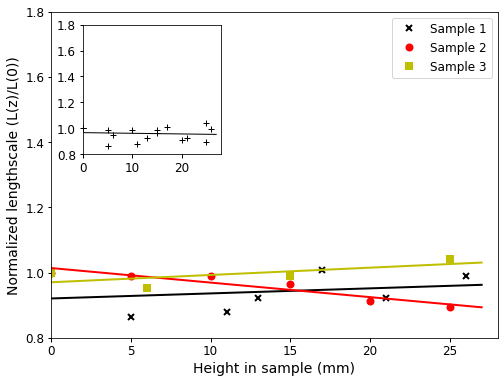}
\caption{Graphs of channel size against height in non-uniform (left) and uniform (right) bijels. The main graphs show different samples in different colours and the inset graphs show all samples collated. The lines are linear least-squares fits for each dataset.}
  \label{fig:LengthscaleGraph}
 \end{center}
\end{figure*}

Figure \ref{fig:LengthscaleGraph} shows how the channel width changes with height in bijel samples, both for samples with a designed gradient and for control samples made using a conventional protocol. The normalised lengthscale is calculated by dividing the channel width by the channel width measured at the bottom of the same sample. The 48 data points for the non-uniform data are taken from 8 separate bijel samples, and the 17 data points for the uniform data are taken from 3 separate bijel samples. Qualitatively, these results suggest that our method to produce non-uniform bijels indeed successfully creates bijels with a gradient in lengthscale.

The coefficient of determination, $R^2$, for a linear line of best fit through the non-uniform data is 0.43, corresponding to a Student's $t$ value of 5.85. We can therefore reject the null hypothesis that the gradient is zero in the case of the non-uniform data, with $p<10^{-6}$ \cite{Howell}. For the uniform data, $R^2=0.008$ and $t=0.35$, and so we cannot reject the null hypothesis that the gradient is zero in this case ($p>0.3$). Notably, $R^2$ is a measure of how much better a fit of the functional form $y=mx+c$ is than a fit of the functional form $y=\bar{y}$. If the slope $m$ is close to 0, then there is little difference between these two fits, which explains why $R^2$ is small for the uniform data i.e.~any variation in the uniform data is due to noise rather than any correlation between the variables. Hence, the statistical analysis confirms that our method is successfully creating bijels with a gradient in lengthscale.

Finally, we note that the gradient in measured lengthscale (Fig.~\ref{fig:LengthscaleGraph} (left)) is lower than that predicted by the gradient in local particle concentration (Fig.~\ref{fig:TransmissionFig}, Eqns.~\ref{Eqn:LocalPhiEqn} and \ref{Eqn:AbsorbanceEqn}). Possible explanations for this include polydispersity in the particle size distribution and Eqn.~\ref{Eqn:BeerLambertEqn} not holding at the particle volume fractions used here; further experiments may provide an explanation.

\section{Conclusions and Outlook}
We have created and characterised bijels which have a gradient in channel size by developing a simple method which creates particle dispersions with a gradient in particle concentration. Since the channel size of a bijel is determined by the local particle concentration, this leads to a gradient in the resultant bijel. We have quantitatively compared these non-uniform bijels to conventional uniform bijels.

We envisage that other methods for creating binary fluids with a gradient in particle concentration will also lead to bijels with a gradient in channel width. Such methods could include the use of polydisperse particle dispersions, which would create a gradient in particle concentration as they sediment, and the agitation of previously-sedimented particle dispersions. Methods for more finely controlling the local particle concentration would also lead to more complex bijel geometries being realised. Finally, external fields may well be promising: Rumble \emph{et al.}~showed that centrifugation can be used to obtain a gradient in channel size, though it also aligns both channels perpendicularly to the direction of the centrifugal force~\cite{Rumble_16}.

Tailored soft structures such as those we have described here could be processed in exactly the same way as those in Refs.~\cite{Lee_13,Witt_16}, in order to create electrodes which have a gradient in channel size. Our method could also be used in conjunction with the methods described in Ref.~\cite{Ching_21} to create bijel-templated materials with a gradient in channel size, using rapid selective polymerisation. Such samples could be more suited to electrochemical applications than conventional, uniform, bijels since they could optimise diffusive transport of ions and storage capacity.

\section{Acknowledgements}
We thank EPSRC for providing funding via Grant EP/P007821/1 and P.S.~Clegg for useful initial discussions.


\end{document}